# Accurate frequency referencing for fieldable dual-comb spectroscopy

**Gar-Wing Truong*, Eleanor M. Waxman, Kevin C. Cossel, Esther Baumann, Andrew Klose, Fabrizio R. Giorgetta, William C. Swann, Nathan R. Newbury, Ian Coddington**

*National Institute of Standards and Technology, 325 Broadway, Boulder, Colorado 80305, USA*
**garwing.truong@nist.gov*

**Abstract:** A fieldable dual-comb spectrometer is described based on a "bootstrapped" frequency referencing scheme in which short-term optical phase coherence between combs is attained by referencing each to a free-running diode laser, whilst high frequency resolution and long-term accuracy is derived from a stable quartz oscillator. This fieldable dual-comb spectrometer was used to measure spectra with full comb-tooth resolution spanning from 140 THz (2.14 µm, 4670 $cm^{-1}$) to 184 THz (1.63 µm, 6140 $cm^{-1}$) in the near infrared with a frequency sampling of 200 MHz (0.0067 $cm^{-1}$), ~ 120 kHz frequency resolution, and ~ 1 MHz frequency accuracy. High resolution spectra of water and carbon dioxide transitions at 1.77 µm, 1.96 µm and 2.06 µm show that the molecular transmission acquired with this fieldable system did not deviate from those measured with a laboratory-based system (referenced to a maser and cavity-stabilized laser) to within $5.6\times10^{-4}$. Additionally, the fieldable system optimized for carbon dioxide quantification at 1.60 µm, demonstrated a sensitivity of 2.8 ppm-km at 1 s integration time, improving to 0.10 ppm-km at 13 minutes of integration time.

## References and links

1. I. Coddington, N. Newbury, and W. Swann, "Dual-comb spectroscopy," Optica **3**, 414 (2016).
2. L. C. Sinclair, J.-D. Deschênes, L. Sonderhouse, W. C. Swann, I. H. Khader, E. Baumann, N. R. Newbury, and I. Coddington, "Invited Article: A compact optically coherent fiber frequency comb," Rev. Sci. Instrum. **86**, 81301 (2015).
3. L. C. Sinclair, I. Coddington, W. C. Swann, G. B. Rieker, A. Hati, K. Iwakuni, and N. R. Newbury, "Operation of an optically coherent frequency comb outside the metrology lab," Opt. Express **22**, 6996–7006 (2014).
4. J. Lee, K. Lee, Y.-S. Jang, H. Jang, S. Han, S.-H. Lee, K.-I. Kang, C.-W. Lim, Y.-J. Kim, and S.-W. Kim, "Testing of a femtosecond pulse laser in outer space," Sci. Rep. **4**, 5134 (2014).
5. M. Lezius, T. Wilken, C. Deutsch, M. Giunta, O. Mandel, A. Thaller, V. Schkolnik, M. Schiemangk, A. Dinkelaker, M. Krutzik, A. Kohfeldt, A. Wicht, A. Peters, O. Hellmig, H. Duncker, K. Sengstock, P. Windpassinger, T. W. Haensch, and R. D. Holzwarth, "Frequency comb metrology in space," in *8th Symposium on Frequency Standards and Metrology* (2015).
6. N. Kuse, J. Jiang, C.-C. Lee, T. R. Schibli, and M. E. Fermann, "All polarization-maintaining Er fiber-based optical frequency combs with nonlinear amplifying loop mirror," Opt. Express **24**, 3095–3102 (2016).
7. P. J. Schroeder, R. J. Wright, S. Coburn, B. Sodergren, K. C. Cossel, S. Droste, G. W. Truong, E. Baumann, F. R. Giorgetta, I. Coddington, N. R. Newbury, and G. B. Rieker, "Dual frequency comb laser absorption spectroscopy in a 16 MW gas turbine exhaust," Proc. Combust. Inst. (2016), http://dx.doi.org/10.1016/j.proci.2016.06.032
8. S. Coburn, R. Wright, K. C. Cossel, G.-W. Truong, E. Baumann, I. Coddington, N. R. Newbury, C. Alden, S. Ghosh, K. Prasad, and G. B. Rieker, "Methane Detection for Oil and Gas Production Sites Using Portable Dual-Comb Spectrometry," in *71st International Symposium on Molecular Spectroscopy* (2016), TB04.
9. G.-W. Truong, E. Waxman, K. C. Cossel, F. R. Giorgetta, W. C. Swann, I. Coddington, and N. R. Newbury, "Dual-comb Spectroscopy for City-scale Open Path Greenhouse Gas Monitoring," in *CLEO 2016* (2016), SW4H.2.
10. A. M. Zolot, F. R. Giorgetta, E. Baumann, J. W. Nicholson, W. C. Swann, I. Coddington, and N. R. Newbury, "Direct-comb molecular spectroscopy with accurate, resolved comb teeth over 43 THz," Opt. Lett. **37**, 638–640 (2012).

11. S. Okubo, K. Iwakuni, H. Inaba, K. Hosaka, A. Onae, H. Sasada, and F.-L. Hong, "Ultra-broadband dual-comb spectroscopy across 1.0–1.9 µm," Appl. Phys. Express **8**, 82402 (2015).
12. P. Giaccari, J.-D. Deschênes, P. Saucier, J. Genest, and P. Tremblay, "Active Fourier-transform spectroscopy combining the direct RF beating of two fiber-based mode-locked lasers with a novel referencing method," Opt. Express **16**, 4347–4365 (2008).
13. J.-D. Deschênes, P. Giaccarri, and J. Genest, "Optical referencing technique with CW lasers as intermediate oscillators for continuous full delay range frequency comb interferometry," Opt. Express **18**, 23358–23370 (2010).
14. J. Roy, J.-D. Deschênes, S. Potvin, and J. Genest, "Continuous real-time correction and averaging for frequency comb interferometry," Opt. Express **20**, 21932–21939 (2012).
15. T. Ideguchi, A. Poisson, G. Guelachvili, T. W. Hänsch, and N. Picqué, "Adaptive dual-comb spectroscopy in the green region," Opt. Lett. **37**, 4847–4849 (2012).
16. T. Ideguchi, A. Poisson, G. Guelachvili, N. Picqué, and T. W. Hänsch, "Adaptive real-time dual-comb spectroscopy," Nat. Commun. **5**, 3375 (2014).
17. M. Cassinerio, A. Gambetta, N. Coluccelli, P. Laporta, and G. Galzerano, "Absolute dual-comb spectroscopy at 1.55 µm by free-running Er:fiber lasers," Appl. Phys. Lett. **104**, 231102 (2014).
18. T. Yasui, R. Ichikawa, Y.-D. Hsieh, K. Hayashi, H. Cahyadi, F. Hindle, Y. Sakaguchi, T. Iwata, Y. Mizutani, H. Yamamoto, K. Minoshima, and H. Inaba, "Adaptive sampling dual terahertz comb spectroscopy using dual free-running femtosecond lasers," Sci. Rep. **5**, 10786 (2015).
19. D. A. Long, A. J. Fleisher, K. O. Douglass, S. E. Maxwell, K. Bielska, J. T. Hodges, and D. F. Plusquellic, "Multiheterodyne spectroscopy with optical frequency combs generated from a continuous-wave laser," Opt. Lett. **39**, 2688–2690 (2014).
20. P. Martín-Mateos, B. Jerez, and P. Acedo, "Dual electro-optic optical frequency combs for multiheterodyne molecular dispersion spectroscopy," Opt. Express **23**, 21149 (2015).
21. G. Millot, S. Pitois, M. Yan, T. Hovhannisyan, A. Bendahmane, T. W. Hänsch, and N. Picqué, "Frequency-agile dual-comb spectroscopy," Nat. Photonics **10**, 27–30 (2016).
22. T. Ideguchi, T. Nakamura, Y. Kobayashi, and K. Goda, "Kerr-lens mode-locked bidirectional dual-comb ring laser for broadband dual-comb spectroscopy," Optica **3**, 748 (2016).
23. X. Zhao, G. Hu, B. Zhao, C. Li, Y. Pan, Y. Liu, T. Yasui, and Z. Zheng, "Picometer-resolution dual-comb spectroscopy with a free-running fiber laser," Opt. Express **24**, 21833 (2016).
24. F. Adler, K. Moutzouris, A. Leitenstorfer, H. Schnatz, B. Lipphardt, G. Grosche, and F. Tauser, "Phase-locked two-branch erbium-doped fiber laser system for long-term precision measurements of optical frequencies," Opt Express **12**, 5872–5880 (2004).
25. G. B. Rieker, F. R. Giorgetta, W. C. Swann, J. Kofler, A. M. Zolot, L. C. Sinclair, E. Baumann, C. Cromer, G. Petron, C. Sweeney, P. P. Tans, I. Coddington, and N. R. Newbury, "Frequency-comb-based remote sensing of greenhouse gases over kilometer air paths," Optica **1**, 290–298 (2014).
26. F. R. Giorgetta, G. B. Rieker, E. Baumann, W. C. Swann, L. C. Sinclair, J. Kofler, I. Coddington, and N. R. Newbury, "Broadband Phase Spectroscopy over Turbulent Air Paths," Phys. Rev. Lett. **115**, 103901 (2015).
27. I. Coddington, W. C. Swann, and N. R. Newbury, "Coherent linear optical sampling at 15 bits of resolution," Opt. Lett. **34**, 2153–2155 (2009).
28. A. M. Zolot, F. R. Giorgetta, E. Baumann, W. C. Swann, I. Coddington, and N. R. Newbury, "Broad-band frequency references in the near-infrared: Accurate dual comb spectroscopy of methane and acetylene," J. Quant. Spectrosc. Radiat. Transf. **118**, 26–39 (2013).
29. T.-A. Liu, R.-H. Shu, and J.-L. Peng, "Semi-automatic, octave-spanning optical frequency counter," Opt Express **16**, 10728–10735 (2008).
30. I. Coddington, W. C. Swann, and N. R. Newbury, "Coherent dual-comb spectroscopy at high signal-to-noise ratio," Phys Rev A **82**, 43817 (2010).
31. N. R. Newbury, I. Coddington, and W. C. Swann, "Sensitivity of coherent dual-comb spectroscopy," Opt. Express **18**, 7929–7945 (2010).
32. L. S. Rothman, I. E. Gordon, A. Barbe, D. C. Benner, P. E. Bernath, M. Birk, V. Boudon, L. R. Brown, A. Campargue, J. P. Champion, K. Chance, L. H. Coudert, V. Dana, V. M. Devi, S. Fally, J. M. Flaud, R. R. Gamache, A. Goldman, D. Jacquemart, I. Kleiner, N. Lacome, W. J. Lafferty, J. Y. Mandin, S. T. Massie, S. N. Mikhailenko, C. E. Miller, N. Moazzen-Ahmadi, O. V. Naumenko, A. V. Nikitin, J. Orphal, V. I. Perevalov, A. Perrin, A. Predoi-Cross, C. P. Rinsland, M. Rotger, M. Simeckova, M. A. H. Smith, K. Sung, S. A. Tashkun, J. Tennyson, R. A. Toth, A. C. Vandaele, and J. Vander Auwera, "The HITRAN 2008 molecular spectroscopic database," J. Quant. Spectrosc. Radiat. Transf. **110**, 533–572 (2009).

## 1. Introduction

Dual-comb spectroscopy (DCS) is emerging as a powerful technique for comb-tooth resolved, broadband spectroscopy with high optical frequency accuracy, rapid update rates and high signal-to-noise ratio (SNR) [1]. Although these aspects of DCS make it an attractive option for remote sensing in field applications, the full quantitative performance of DCS has only been demonstrated with stabilized frequency combs phase-locked to laboratory references. Recent advances in fiber-based frequency combs can now provide broadband near-infrared light in relatively robust packages that are capable of operation outside laboratory environments [2–6]. However, an accurate DCS instrument requires: i) sub-radian mutual optical coherence between the combs to achieve high SNR, comb-tooth resolved spectra, ii) absolute comb-tooth linewidth that is much less than the desired spectral resolution (which is not guaranteed solely by condition (i)), and iii) absolute comb-tooth frequency accuracy to avoid the need for separate spectral frequency calibration. Here we demonstrate such a DCS frequency stabilization scheme that meets these challenges and is compatible with field operation. Although, the DCS system could be further engineered for compactness, this configuration has already supported recent field measurements at a 16 MW gas turbine [7] as well as city-scale open-path operation outside a laboratory [8,9].

In the near infrared, one way to establish mutual comb coherence, long-term stability, and frequency accuracy is to phase lock the mode-locked laser-based frequency combs to a common cavity-stabilized continuous-wave (CW) laser [10,11]. However, the reliance on a stable optical cavity reference is an obvious downside in terms of cost, portability and robustness. Several approaches remove the need for a cavity reference by using free-running combs and applying digital [12–14] and analog [15–18] corrections based on the measured differential phase noise on a comb tooth. This effectively establishes a mutual coherence between the combs without the need for the stable cavity and can in some cases provide sufficient mutual coherence for extremely high SNR operation [14]. Modulator-based combs and common-comb-cavity designs have also been employed to similar effect but with narrower optical bandwidth [19–23]. However, the downside of these free-running approaches is that while the spectra may be resolved to individual comb teeth, the comb teeth themselves are not fixed but can undergo large frequency excursions in noisy environments, sacrificing the intrinsic spectral resolution possible with DCS. Furthermore, there is no absolute knowledge of the frequency axis, so additional information, such as known spectral lines or wavemeter measurements, must be used to re-establish absolute optical frequencies at some relatively coarse level compared to self-referenced laboratory frequency combs.

Here we implement a fieldable frequency referencing scheme that maintains comb coherence as well as long-term frequency stability, providing high spectral resolution and accuracy. The system is based on two self-referenced Er fiber combs [2,3] phase locked to a common, free-running, commercial external cavity diode laser (ECDL) to establish sub-radian mutual optical coherence. The absolute linewidth (i.e. spectral resolution) and frequency accuracy are restored by comparing the comb repetition rate to a commercial ovenized quartz oscillator, and then using that comparison to stabilize the CW laser frequency in a "bootstrapping" operation that takes full advantage of the self-referenced comb. We demonstrate a frequency resolution and accuracy of the spectra at the ~120 kHz and ~ 1 MHz level, respectively, and traceable to the quartz oscillator. Near infrared spectra are acquired spanning 44 THz, consisting of ~220,000 comb-resolved teeth at 200 MHz spacing, with no degradation in resolution or SNR as compared to stabilization with a cavity-stabilized CW laser. Such a DCS system is of general interest for high-accuracy quantitative spectroscopy simultaneously over multiple gas species. An immediate application would be in atmospheric greenhouse gas sensing. For $CO_2$ detection in particular, stronger absorption cross-sections provide a sensitivity advantage in the ~ 2 µm region. We demonstrate operation from 1.6 µm out to 2.14 µm which enables quantitative comparisons of fieldable and laboratory systems using the 20013$\leftarrow$00001 and 20011$\leftarrow$00001 $CO_2$ absorption bands at 2.06 µm and 1.96 µm, respectively, and selected water transitions at 1.77 µm. A $CO_2$ sensitivity of 2.8 ppm-km was achieved at 1 s and 0.10 ppm-km at 13 minutes. The data verifies that the field-mode configuration maintains the full unique combination of features possible with DCS, including frequency-comb tooth resolution, absolute frequency accuracy, rapid single-spectrum acquisition, high SNR through coherent averaging, and broadband, flexible spectral coverage.

## 2. Design for a fieldable DCS instrument

The DCS instrument uses two Er fiber frequency combs based on a robust, all polarization-maintaining (PM) fiber design described in [2,3]. The comb repetition rates are $f_r \approx 200$ MHz and typically differ by $\Delta f_r$ = 208.3 Hz. A schematic of the setup is shown in Figure 1(a). The output of each oscillator is evenly split into two branches [24], which are independently amplified with erbium-doped fiber amplifiers (EDFAs) and then spectrally broadened with a section of highly non-linear fiber (HNLF). The output of one branch is optimized to produce an octave-spanning spectrum for detection of the carrier-envelope offset (CEO) frequency using *f-2f* interferometry. The output of the second branch generates the broadband light used for spectroscopy. A long-wave-pass optical filter selected optical frequencies from the cutoff frequency of 184 THz (1.63 µm, 6140 cm$^{-1}$) down to the edge of the comb spectrum at 140 THz (2.14 µm, 4670 cm$^{-1}$). This band is of particular utility for monitoring of greenhouse gases, such as carbon dioxide ($CO_2$), methane ($CH_4$) and water ($H_2O$).

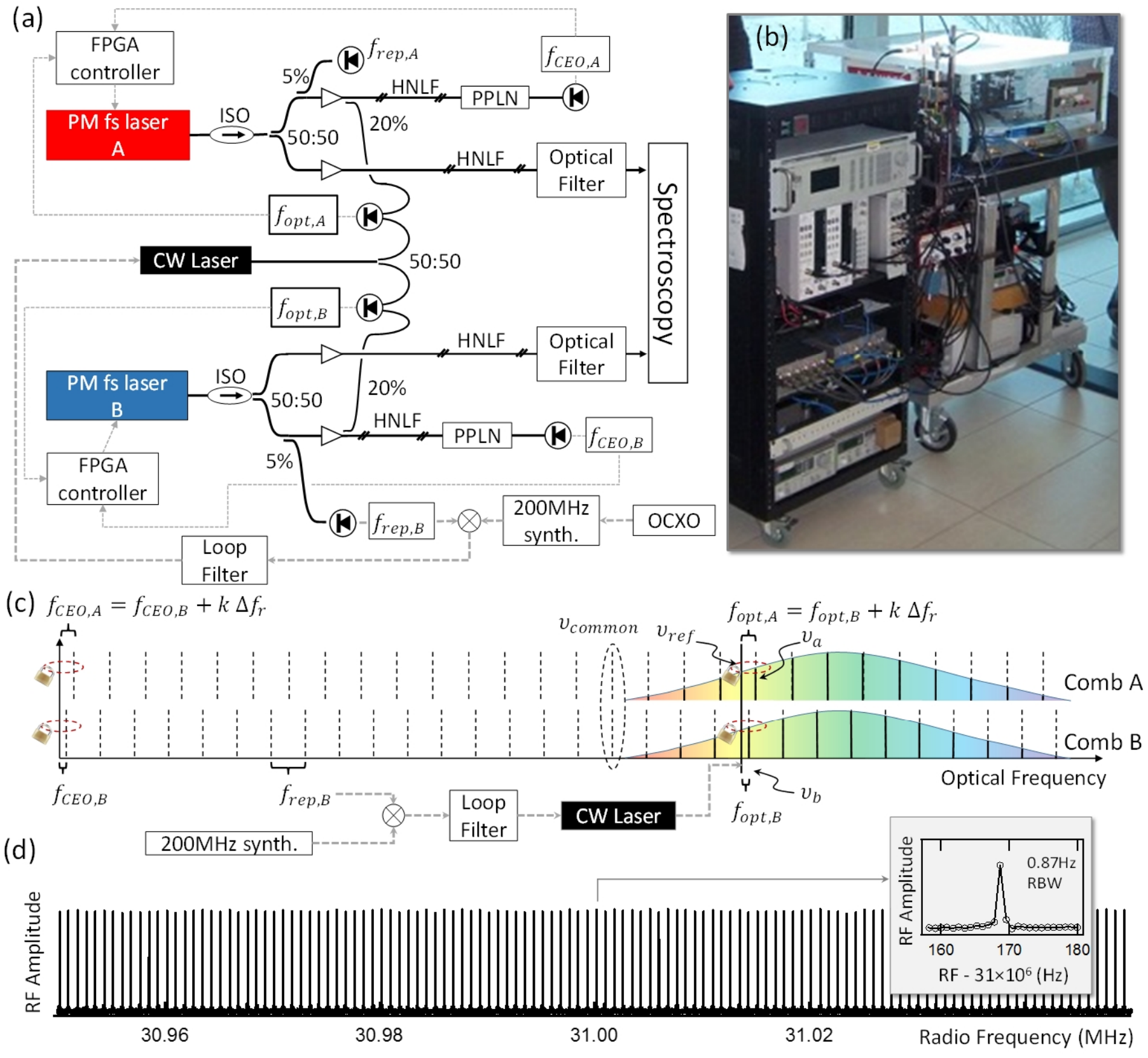

Figure 1. (a) Schematic of the DCS instrument. (b) Photograph of the instrument. The optical components shown in (a) fit within the volume enclosed by the acrylic shield on the equipment trolley. (c) Frequency domain view of the field-compatible DCS phase-locking configuration, as described in the text. (d) A continuously streamed and unapodized DCS signal obtained during field-mode operation showing clearly isolated comb teeth. The inset verifies the sub-Hz mutual coherence between the two frequency combs. RBW: resolution bandwidth. FPGA: field programmable gate array; OCXO: oven controlled crystal quartz oscillator; PPLN: periodically poled lithium niobate; PM: polarization-maintaining; other symbols defined in text.

A schematic of the fieldable DCS instrument is shown in Figure 1(a) and the full system is pictured in Figure 1(b). The optical components fit in a 70 x 65 x 25 cm$^3$ volume, making the system transportable on an equipment cart. Similarly, the pump lasers and auxiliary electronics fit in a mobile 19-inch equipment rack. The fiber-coupled spectroscopic output provides flexibility in delivering the light to a variety of optical paths for gas measurement, e.g. coupling to a multipass gas cell for laboratory-based measurements, or to a telescope for sampling across an outdoor path [8,9,25,26].

### *2.1 Phase-locking configuration for fieldable operation*

As shown in Figure 1(c), the two combs are defined by their respective phase-locked carrier-envelope offset frequencies, $f_{CEO,A}$ and $f_{CEO,B}$ and their phase-locked optical comb tooth frequencies $\upsilon_A$ and $\upsilon_B$. The optical frequencies $\upsilon_A$ and $\upsilon_B$ are defined by the sum of the optical reference frequency, $\upsilon_{ref}$, and the rf phase-locking frequencies $f_{opt,A}$ and $f_{opt,B}$, as shown in Figure 1(c). This locking configuration results in phase-coherent interferograms that are exactly periodic at $\Delta f_r$, permitting simple real-time continuous signal co-adding for long-term averaging and high SNR [27,28].

The reference laser is a commercial ECDL at 1560.62nm. The ECDL is packaged in a robust and portable 14-pin butterfly package with a fiber-optic output and has an instantaneous Lorentzian linewidth < 1 kHz. A digital phase lock ensures both optical phase locks have the same loop parameters and therefore mutually "follow" the phase of the ECDL up to the lock bandwidth of ~ 100 kHz with < 1 rad of total residual phase error. As shown by the inset in Figure 1(c), this mutual coherence supports time-bandwidth-limited linewidths of 0.87 Hz in the DCS spectrum. However, since the ECDL's frequency will drift, the resolution and frequency accuracy of the spectra will be degraded. We use self-referenced locking of the comb to re-establish absolute accuracy and stability while maintaining the co-adding condition as follows. First, a common rf frequency is used to phase lock the CEO and the optical comb tooth for each comb, i.e. $f_{CEO,A} = f_{opt,A}$ and $f_{CEO,B} = f_{opt,B}$. In that case, the repetition frequency of each comb is exactly related to the phase locked comb tooth as $f_{rep,A} = \upsilon_{ref} / a$ and $f_{rep,A} = \upsilon_{ref} / b$, where $a$ and $b$ are the mode numbers corresponding to the modes at $\upsilon_A$ and $\upsilon_B$. By measuring these repetition rates against the ovenized quartz oscillator, we can rely on well-established dual-comb Vernier techniques [29] to determine the CW laser frequency $\upsilon_{ref}$. Finally, to then stabilize this reference laser frequency to the ovenized quartz oscillator, we phase lock $f_{rep,B}$ against a quartz-referenced rf synthesizer by tuning the center frequency of the ECDL with a feedback bandwidth of less than 100 Hz to ensure long-term frequency stability. The value of $\upsilon_{common}$, the optical frequency that is mapped to 0 Hz in the rf, can be chosen so that it does not fall within the spectrum of interest by offsetting $f_{CEO,A}$ and $f_{opt,A}$ by an integer multiple, *k*, of the repetition rate difference from the frequencies $f_{CEO,B}$ and $f_{opt,B}$ (as shown in Figure 1(c)).

The absolute comb linewidth at 1560 nm was 120 kHz (full-width half-maximum), as measured against the narrow cavity-stabilized laser. This linewidth sets the spectral resolution, which is far below the several-GHz-wide atmospheric pressure-broadened molecular transitions. Furthermore, the fractional absolute frequency accuracy of each comb tooth (and ultimately the DCS spectrum) matches the quartz oscillator's accuracy, which is $6\times10^{-9}$ in our setup. If the quartz oscillator is additionally GPS-disciplined, the relative frequency accuracy improves to $1\times10^{-11}$.

## 3. Experimental measurements and results

### *3.1 Experimental setup*

Figure 2 shows the experimental setup to evaluate the fieldable DCS instrument and to compare its operation in the field-mode with the more conventional lab-mode (i.e. referenced to a cavity-stabilized laser of known frequency). For simplicity, the DCS instrument was operated in the collinear configuration [1], where both comb outputs were combined and launched through the gas. However, to allow for balanced detection and mitigation of relative intensity noise (RIN), the combs were orthogonally polarized before being combined and transmitted through the 30-m multipass Herriott cell. The multi-pass cell was filled with 1 atmosphere of room air (84 kPa), but with a $CO_2$ content enhanced by a factor of 21 to ~8400 ppm (corresponding to the absorbance expected over 630 m of air). Balanced detection was implemented via a halfwave plate and polarizing beamsplitter. A total of 190 µW of the transmitted comb light was incident on a custom free-space, 90 MHz bandwidth, amplified InGaAs balanced detector. When combined with a transimpedance amplifier, the noise-equivalent power was 8 pW/√Hz.

The balanced detector output was digitized synchronously at the repetition rate of one comb, with interferograms occurring at the repetition rate difference $\Delta f_r$=208.3 Hz. Real-time coherent averaging in the time domain is performed by simple co-addition of 80 successive interferograms [27,30]. Further co-adding was performed by a software phase correction on the averaged interferograms, thus extending the coherent averaging time to the desired measurement interval.

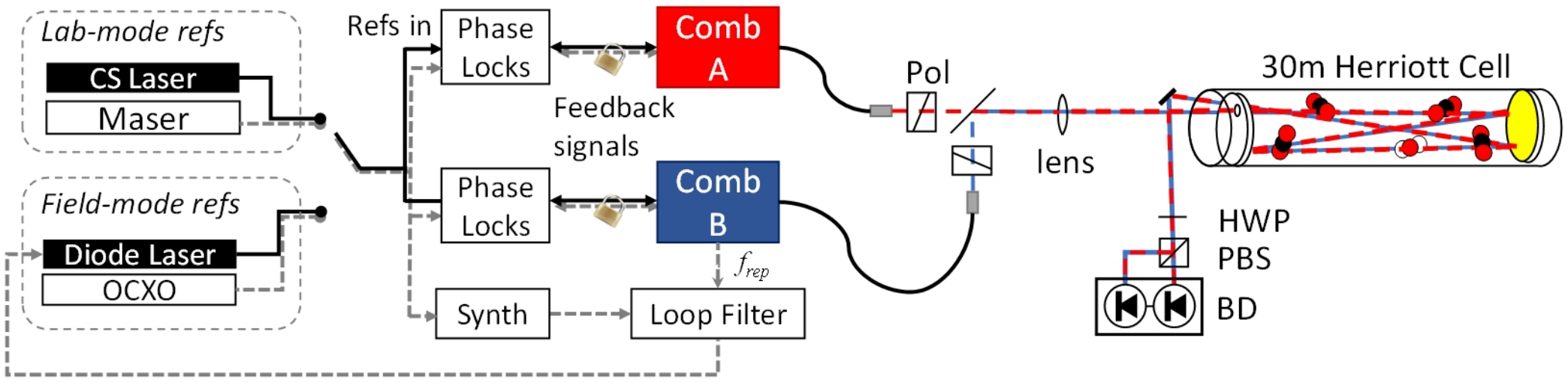

Figure 2. Schematic of the spectroscopy apparatus. Frequency stabilization of the DCS instrument is switched between either a set of lab-based or fieldable frequency references. The Herriott cell contains a fixed sample of $CO_2$-enhanced room air. CS Laser: cavity-stabilized laser; Pol: polarizer; PBS: polarizing beamsplitter; HWP: half-wave plate; BD: balanced detector. Synth: rf synthesizer; OCXO: oven controlled crystal quartz oscillator.

As shown in Figure 2, the DCS instrument could be operated in field-mode using the free-running CW laser and quartz oscillator (as described in Section 2.1) or in lab-mode. The latter replaces the quartz oscillator with a hydrogen maser and the ECDL with a cavity-stabilized laser with a known absolute optical frequency of 192.097 679 THz (1560.625 09 nm) and an instantaneous linewidth of 1 Hz [28]. When switching between lab and field operation, small changes in the repetition rate and $\Delta f_r$ are made to retain the same mode numbers *a* and *b* despite the slightly differing wavelengths of the cavity-stabilized and ECDL CW laser sources.

### *3.2 Results for broadband operation*

The cell transmission was measured continuously over a period of 4 hours in each of the field- and lab- modes of operation. The comb spectrum, 44 THz wide, consisting of ~220,000 comb-resolved teeth, is shown in Figure 3(a). To isolate the gas absorbance from the temporally varying comb spectrum, the incident comb spectral shape must be normalized out either through a reference spectrum or by a polynomial fit, as described in Ref. [25]. The absorbance can then be fit to retrieve the gas concentration. Here the reference spectrum introduced additional etalons and a polynomial spectral normalization was preferred for these narrow absorption lines from small gas molecules.

Figure 3(b-d) shows transmission spectra at 1.77 μm, 1.96 μm and 2.06 μm for both operating modes. Further expanded views of some of the strongest $CO_2$ lines are shown in panels (e-f) and some distinct water lines in (g); it is evident that the spectra measured in the field-mode and lab-mode of operation are identical and no lineshape perturbation due to inaccuracies in the reconstruction of the optical frequency axis has occurred. As shown in Figure 3(h-j), the difference for the full 4-hour data run is generally at or below 1 part in a thousand, limited by a residual etalon-like structure. The spectral SNR at 4-minute averaging time is in the range of 225-1000, improving inversely with the square root of integration time out to ~30 minutes where it begins to be limited by the residual spectral etalon-like structure. Following Refs [1,31], we define a figure of merit as the product of the average spectral SNR and number of spectral elements, yielding an average figure of merit of ~ $1.7\times10^7$ √Hz over the full 44 THz spectrum.

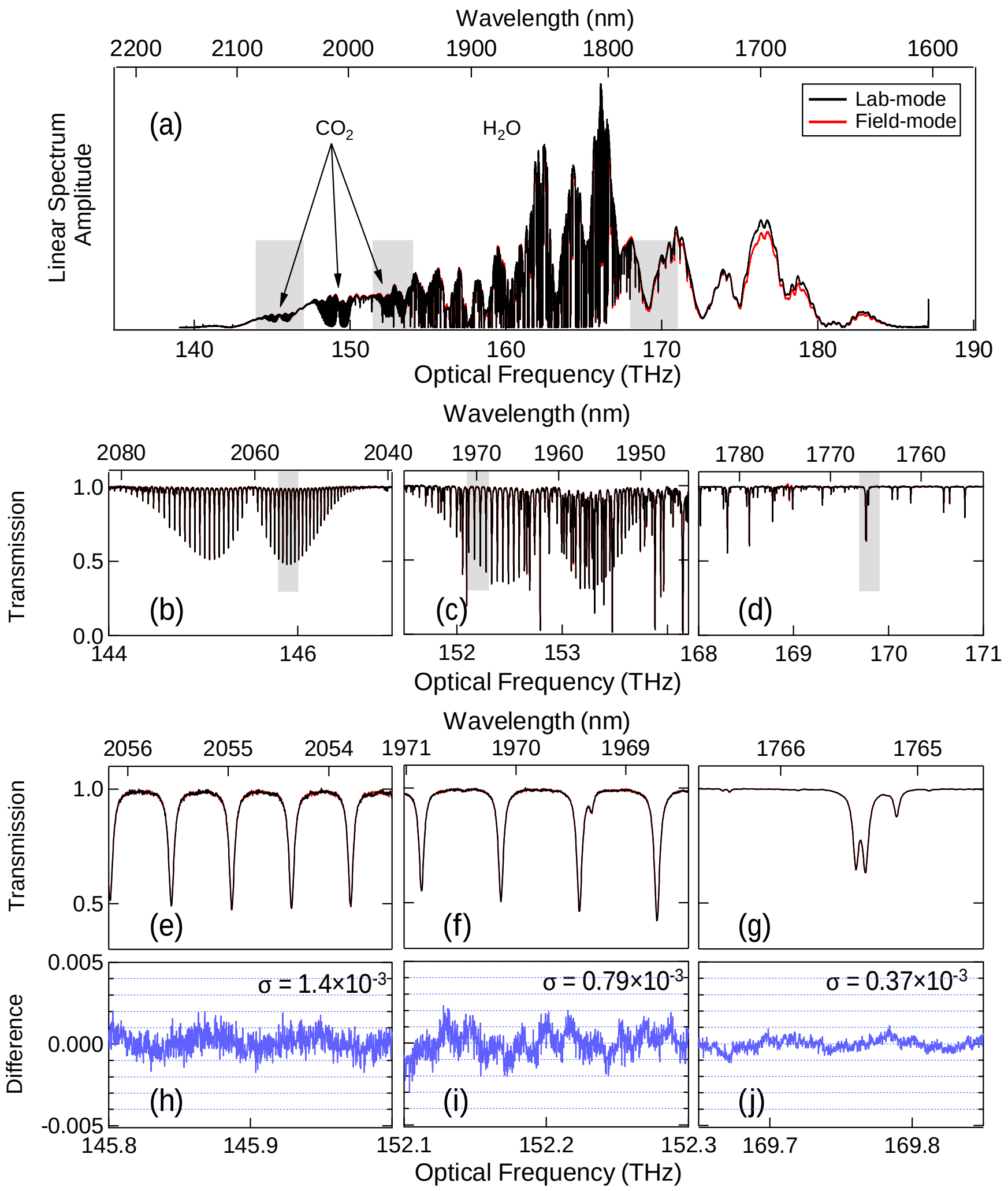

Figure 3. Measured spectra. (a) Broadband optical spectrum measured in lab-mode (black traces) and field-mode (red traces) operation for a 30 m multipass cell filled with $CO_2$-enhanced room air. (b,c,d) Transmission spectra for the three shaded regions of (a) that are dominated, respectively by $CO_2$, $CO_2$ and $H_2O$ lines. (e,f,g) An expanded view of the shaded regions in (b), (c) and (d), respectively, for the lab-mode, field-mode and for their difference after 4 hours of averaging (blue). (h,i,j) The rms differences between lab- and field-modes of operation are shown. In panels (b-g), the red and black traces are nearly identical.

Figure 4(a) is a further expanded view of the P(24) line from the 20011←00001 band showing that the rms difference between spectra measured in the two modes of operation is $5.6\times10^{-4}$ at 200 MHz spacing. Any inaccuracies in the reconstruction of the optical axis in the fieldable configuration would lead to a frequency shift or distortion of the lineshape that would lead to a distinctive trend in the spectral difference, which is not observed. In addition, Figure 4(b) shows the difference in the fitted line centers between the spectra recorded in the laboratory and field configurations for the 45 most intense $CO_2$ lines. The weighted mean was 1.2 MHz, which is negligible compared to the air-broadened linewidths and is similar to the line center uncertainty in the HITRAN database for the strongest $CO_2$ lines [32].

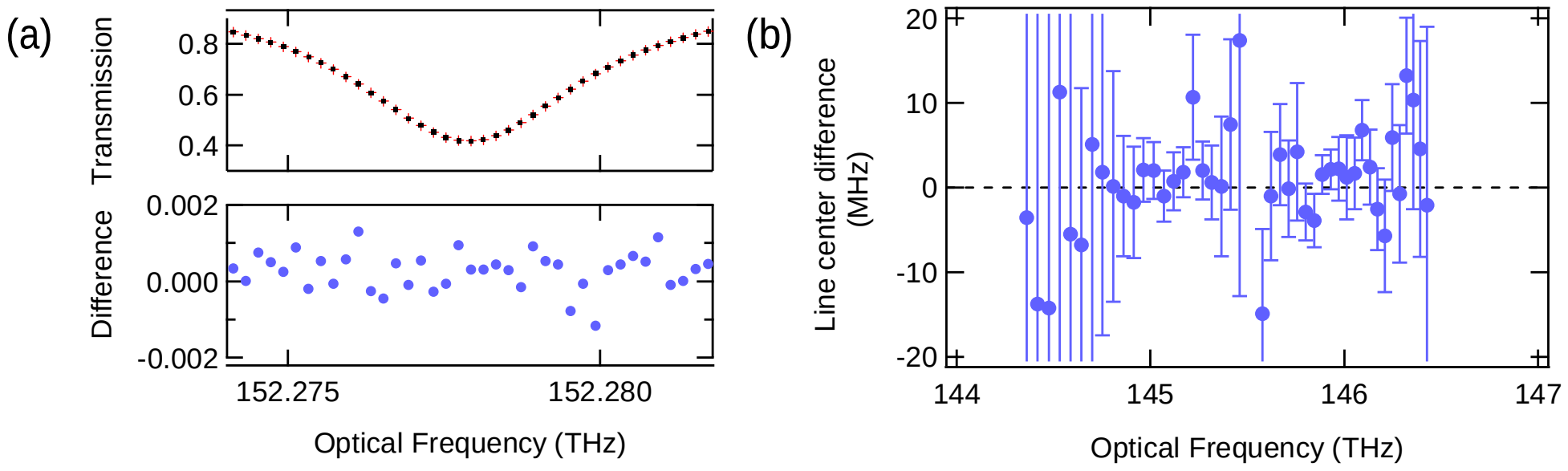

Figure 4. (a) Comparison of the P(24) line of $CO_2$ measured with the lab-mode (black dots) and field-mode (red crosses) operation, sampled at 200.039 967 443 5 MHz intervals set by the mean repetition rate of the two combs. The rms difference between spectra is $5.6\times10^{-4}$ across this span. (b) Comparison of measured line centers for 45 $CO_2$ lines under lab- and field-mode operation. Error bars indicate the statistical uncertainty estimated from least-squares regression and are correlated with the optical depth of each transition.

## 4. Carbon dioxide detection limits

The gas-concentration measurement precision is an important parameter for atmospheric gas monitoring. The gas concentrations were retrieved by a least-squares fit of the spectra around the two non-saturated $CO_2$ bands at 1.96 µm and 2.06 µm using the HITRAN 2008 database [32] . The fit returns the column density, which is converted to mixing ratio units of ppm by normalizing to the 83.99 kPa air pressure in the cell. In the field-mode, the measured $CO_2$ concentration was 8479 ± 12 ppm and 8422 ± 7 ppm, from fitting the 1.96 µm and 2.06 µm bands respectively over a 2500-second window. A consecutive measurement in the lab-mode configuration resulted in 8469 ± 7 ppm and 8427 ± 8 ppm, or identical results for each band to within the uncertainty. The disagreement between spectral bands is attributed to linestrength discrepancies in HITRAN.

To determine the precision of path-integrated $CO_2$ concentration versus time, the spectra were processed at 38-second time intervals and the retrieved concentration was normalized by the path length for units of ppm-km. As shown in Figure 5, in all cases and for both 2 µm $CO_2$ bands, we achieve a sensitivity of 7.8 ppm-km/$\sqrt{\tau}$, where $\tau$ is the averaging time. At 30 minutes the precision reaches a floor of ~ 0.2 ppm-km. This floor is likely limited by the same residual structure observed in Figure 3(h,i). Crucially, Figure 5 shows that the fieldable system did not suffer from any additional drifts beyond those of the lab-mode operation that would limit sensitivity.

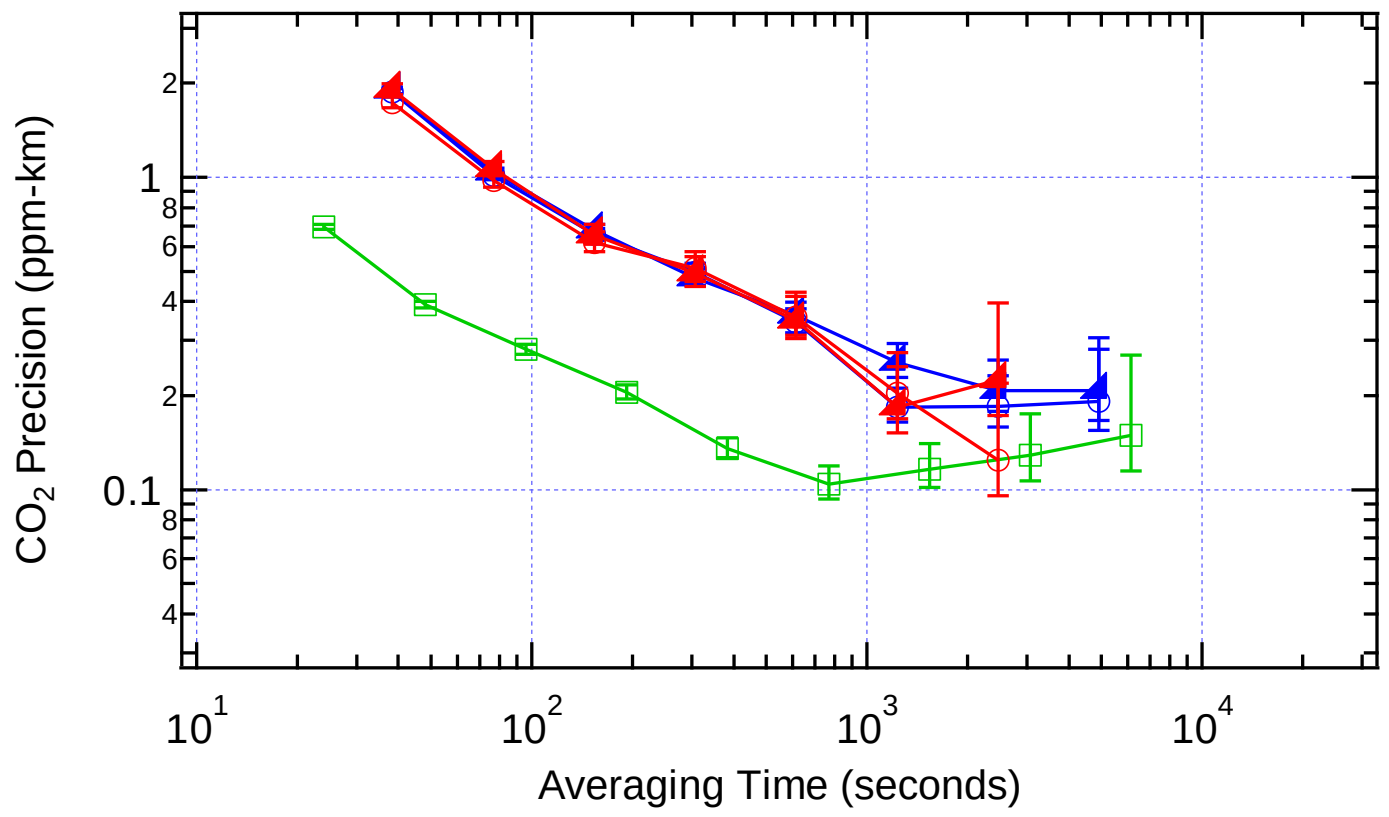

Figure 5. Precision of the $CO_2$ concentration retrievals, as quantified by the Allan deviation in both field-mode (red traces) and lab-modes of operation (blue traces) for the 1.96 µm (circles) and 2.06 µm bands (triangles). In addition, the superior precision for the narrowband configuration targeting $CO_2$ at 1.6 µm is shown (green squares).

Although broadband dual-comb spectra can support detection of multiple gases, the supercontinuum approaches used to generate this bandwidth also generated excess RIN. Furthermore, it is difficult to achieve perfectly balanced detection (and therefore RIN suppression) across such broad optical bandwidths. This suggests that a more spectrally tailored, narrowband approach may be advantageous for the highest sensitivity for a few specific target molecules. To this end, we shortened the HNLF that broaden the fs laser output (Figure 1) to generate a narrower spectrum with a RIN of -143 dB/√Hz, 10 dB lower than for the broadband spectrum. This output was then spectrally filtered to a band from 180.5 THz (1.661 µm) to 191.8 THz (1.563 µm), targeted to $CO_2$ and $CH_4$ detection in the 1.6 µm transparency window of the atmosphere. With this lower RIN and superior balancing, at the 190µW of total comb optical power (limited by photoreceiver saturation), the noise floor is approximately equally limited by detector and shot noise. Figure 5 shows the resulting three times improvement sensitivity to $CO_2$ that is commensurate with the improvement in SNR. Our achieved sensitivity was 2.8 ppm-km at 1 second, dropping to 0.10 ppm-km after 13 minutes.

## 5. Conclusions

In recent years, the advent of robust fiber frequency combs has sparked the evolution of dual-comb spectroscopy from laboratory-based demonstrations towards fieldable systems. In this paper, we describe a fieldable scheme for a DCS instrument to operate independently of laboratory-based rf and optical frequency references but is nevertheless capable of ultra-high spectral resolution, high SNR, and frequency-accurate spectral measurements over 44 THz in the near-IR. Direct comparisons of molecular spectra taken in this field-mode versus lab-mode demonstrate spectra that were identical to within $5.4\times10^{-4}$ and sensitivity to retrieved $CO_2$ path-averaged densities with a precision of 2.8 ppm-km at 1 s and 0.10 ppm-km at 13 minutes. This opens up the possibility of DCS instruments with laboratory-levels of performance in field locations of interest to trace gas detection. An application of topical interest would be high sensitivity and accuracy, path-integrated measurements of $CO_2$ for the quantification and modeling of greenhouse gas fluxes, as well as monitoring of $CH_4$ and other gases across oil/gas fields or storage facilities.

## 6. Acknowledgements

The authors thank I.H. Khader, A.J. Fleisher and G.C. Ycas. This work was supported under the DARPA DSO SCOUT program, ARPA-E MONITOR program, and the NIST Greenhouse Gas and Climate Science Initiative. EMW and KCC are supported by NRC Postdoctoral Fellowships.